\documentclass[3p,times]{elsarticle}

\usepackage{amsmath}

\usepackage{amssymb}

\begin{document}

\begin{frontmatter}




\title{Toward an automated analysis of slow ions in \\ nuclear track emulsion}


\author[a]{K.Z.~Mamatkulov} 
\author[a]{R.R.~Kattabekov}
\author[b]{I.~Ambrozova}
\author[a,c]{D.A.~Artemenkov}
\author[a]{V.~Bradnova}
\author[a]{D.V.~Kamanin}
\author[b]{L.~Majling}
\author[d]{A.~Marey}
\author[b]{O.~Ploc}
\author[a]{V.V.~Rusakova}
\author[e]{R.~Stanoeva}
\author[b]{K.~Turek}
\author[a,f]{A.A.~Zaitsev}
\author[a,f]{P.I.~Zarubin\corref{cor1}}
\author[a,f]{I.G.~Zarubina}

\address[a]{Joint Institute for Nuclear Research, Joliot-Curie 6, Dubna, Moscow region,141980, Russia}
\address[b]{Nuclear Physics Institute, of the ASCR, v.v.i., \u{R}e\u{z} 130, 250 68 \u{R}e\u{z}, Czech Republic}
\address[c]{National Research Nuclear University MEPhI (Moscow Engineering Physics Institute), Kashirskoe Shosse 31,Moscow, 115409, Russia}
\address[d]{Minoufia University, Tala't street, EL azhar territory, Menouf city, Menofia, Egypt}
\address[e]{South-West University, Ivan Michailov st. 66, 2700 Blagoevgrad, Bulgaria}
\address[f]{P. N. Lebedev Physical Institute of the Russian Academy of Sciences, Leninskij Prospekt 53, Moscow, 119991, Russia}

\begin{abstract}
Application of the nuclear track emulsion technique (NTE) in radioactivity and nuclear fission studies is discussed. It is suggested to use a HSP-1000 automated microscope for searching for a collinear cluster tri-partition of heavy nuclei implanted in NTE. Calibrations of $\alpha $-particles and ion ranges in a novel NTE are carried out. Surface exposures of NTE samples to a ${}^{252}$Cf source started. Planar events containing fragments and long-range $\alpha $-particles as well as fragment triples only are studied. NTE samples are calibrated by ions Kr and Xe of energy of 1.2 and 3 \textit{A} MeV.
\end{abstract}

\cortext[cor1]{Corresponding author. Tel.: +7-496-216-3403; fax: +7-496-216-5180.\\
\textit{E-mail address:} zarubin@lhe.jinr.ru}
\end{frontmatter}




\vspace*{-8pt}
\section{Introduction}

Nuclear track emulsion (NTE) retains the status of a universal and inexpensive detector in spite of the fact that half a century passed since its development. With unsurpassed spatial resolution of NÒE provides complete observation of tracks starting from fission fragments and down to relativistic particles [1-3]. Unique opportunities of NTE deserve further use in fundamental and applied research in state-of-art accelerators and reactors, as well as with sources of radioactivity, including natural ones. Application of NTE is especially justified in those pioneering experiments in which nuclear particle tracks cannot be reconstructed with the help of electronic detectors.

The NTE technique is based on intelligence, vision and performance of researchers using traditional microscopes. Despite widespread interest, its labor consumption causes limited samplings of hundreds of measured tracks which present as a rule only tiny fractions of the available statistics. Implementation of computerized and fully automated microscopes in the NTE analysis allows one to bridge this gap. These are complicated and expensive devices of collective or even remote usage allow one to describe unprecedented statistics of short nuclear tracks.

To make such a development purposeful it is necessary to focus on such a topical issue of nuclear physics the solution of which can be reduced to simple tasks of recognition and measurement of tracks in NTE to be solved with the aid of already developed programs. One of the suggested problems is a search for the possibility of a collinear cluster tri-partition [4]. The existence of this phenomenon could be established in observations of such a type of ternary fission of heavy nuclei in which a lightest fragment is emitted in the direction of one of the heavy fragments.

Despite distinct observability of fission fragments they can not be fully identified in NTE. However, NTE is valuable due to combination of the best angular resolution and maximum sensitivity. Besides, it is possible to measure the lengths and thicknesses of tracks, and, thus, to classify the fragments. As an initial stage, to provide statistics of ternary fissions it is suggested to analyze a sufficient NTE area exposed to ${}^{252}$Cf source with an appropriate density of tracks of $\alpha $-particles and spontaneous fission fragments. Such an approach will be developed by a NTE with an admixture of the ${}^{252}$Cf isotope [5,6]. Another option is exposure of NTE manufactured with a ${}^{235}$U isotope addition by thermal neutrons.

A large-scale NTE scanning is suggested to be performed on the microscope HSP-1000 [7] of the Department of radiation dosimetry (DRD) of Nuclear Physics Institute of the ASCR, v. v. i.. The use of the NTE resolution will be full if the microscope will be adapted to operate with lenses of the highest magnification. Development of algorithms for automatic search and analysis of short tracks of heavy ions in NTE will be required. On the experimental side, ion ranges in NTE must be calibrated in the $\alpha $-decay and fission energy scale. Progress of the preparatory phase of the proposed study is summarized below.

\vspace*{-8pt}
\section{Calibration by ${\alpha}$-particles}

Production of NTE of the BR-2 type possessing sensitivity to relativistic particles lasted in Moscow for four decades and ended about ten years ago. The interest in its further application stimulated the production of NTE in the MICRON workshop that is part of the company "Slavich" (Pereslavl Zalessky) [8]. At present, NTE samples are produced by layers of thickness of 50 to 200 $\mu $m on glass substrates. Verification of the reproduced NTE in exposures to relativistic particles confirmed that it is similar to the BR-2 NTE. It was decided to demonstrate that NTE is competitive in experiments involving measurements of $\alpha $-particle and heavy ion tracks on a KSM microscope with a 90$\times$ objective.

Correlation of $\alpha $-particle triples were studied in disintegrations of carbon nuclei of NTE composition by 14.1 MeV neutrons [9]. When measuring decays of ${}^{8}$He nuclei implanted in NTE the possibilities of $\alpha $-spectrometry were verified and the effect of the ${}^{8}$He atom drift was established [10-12]. The angular correlations of ${}^{7}$Li and ${}^{4}$He nuclei produced in disintegrations of boron nuclei by thermal neutrons \textit{n}${}_{th}$ were studied in boron enriched NTE[13]. In the last case a mean Li range (at RMS) is equal to 3.1 $\pm$ 0.3 (0.8) $\mu$m at a mean thickness of 0.73 $\pm$ 0.02 (0.05) $\mu$m, and the ${}^{4}$He one is 5.5 $\pm$ 0.5 (1.1) $\mu$m and 0.53 $\pm$ 0.01 (0.04) $\mu$m, respectively. In this series of exposures, the angular resolution of NTE was confirmed to be perfect by expected physical effects, which are manifested in the distributions of the opening angles of the products of the studied reactions.

Surface exposures of NTE samples in DRD were performed by a manually moving ${}^{252}$Cf source. Most likely, the ${}^{252}$Cf isotope decays by emission of $\alpha $-particles of energy of 5-6 MeV, the tracks of which mainly populate an exposed sample. This isotope also undergoes a spontaneous fission to a pair or even triple of fragments with probabilities of 3\%, and about 0.1\%, respectively. For comparison an NTE sample was exposed to a ${}^{241}$Am source emitting only $\alpha $-particles in the same energy range. Since the ranges of decay products are small the source exposures are performed without a light protective paper in a darkroom when illuminated with red light.

In the case of a surface exposure there should not be observed more than two ternary fission fragments as the third one is emitted in the contacting source side. The sign of a ${}^{252}$Cf exposure consists in presence of $\alpha $-particle tracks from ternary fission events whose ranges significantly exceed the decay $\alpha $-particle ranges. This channel dominates in the ${}^{252}$Cf ternary fission having a 90\% probability. Fig. 1 summarizes the measured $\alpha $-particle ranges in the exposures listed above as well as their energy values on the inset are calculated in the SRIM model [14]. Average values of ranges and energy are given in Table 1.

\begin{figure}\vspace*{4pt}
\centerline{\includegraphics*[width=6.5in, height=3.23in, keepaspectratio=false]{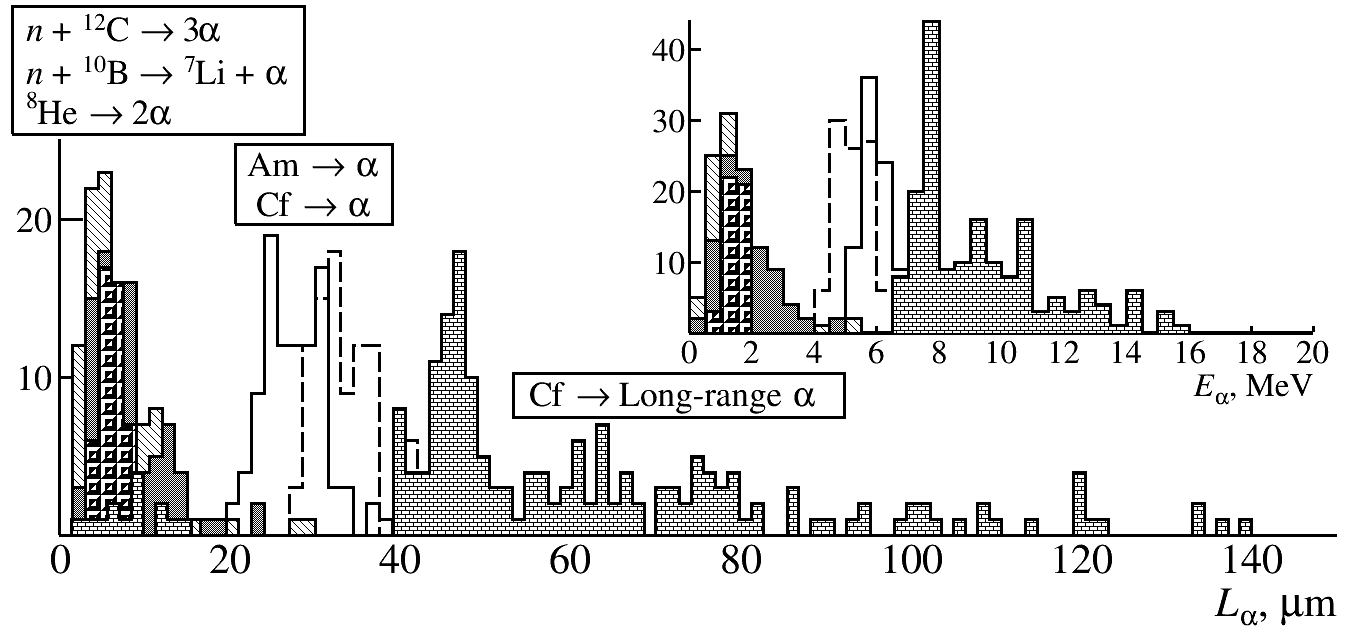}}
\caption{Distributions of $\alpha $-particle ranges: \textit{n}(14.1 MeV) + ${}^{12}$C $\rightarrow$ 3$\alpha$ (obliquely-shaded), ${}^{8}$He $\rightarrow$ 2$\alpha $ (gray), \textit{n}${}_{th}$ + ${}^{10}$B $\rightarrow$ ${}^{7}$Li + $\alpha $ (black dot), Cf $\rightarrow$ $\alpha $ (solid), Am $\rightarrow$ $\alpha $ (dotted histogram), Cf $\rightarrow$ long-range $\alpha $ (brick-shaded); the inset: corresponding of $\alpha $-particle energy estimated via spline-interpolation of range-energy calculations in the SRIM model.}
\end{figure}
\vspace*{-8pt}

\begin{table}
\caption{Average values of ranges and energy of $\alpha$-particles in the studied reactions and decays; values in parentheses are the RMS.}
\begin{center}\begin{tabular}{c c c} \hline 
Reaction or decay & Average range, $\mu$m & Average energy, MeV \\ \hline 
\textit{n}(14.1 MeV) + ${}^{12}$C $\rightarrow$ 3$\alpha $ & 5.8 $\pm$ 0.2 (3.3) & 1.9 $\pm$ 0.05 (0.9) \\ 
${}^{8}$He (2$\beta $) $\rightarrow$ 2$\alpha $ & 7.4 $\pm$ 0.2 (3.8) & 1.7 $\pm$ 0.03 (0.8) \\ 
\textit{n}${}_{th}$+${}^{10}$B $\rightarrow$ ${}^{7}$Li + $\alpha $ & 5.5 $\pm$ 0.5 (1.1) & 1.4 $\pm$ 0.5 (0.3) \\ 
${}^{241}$Am $\rightarrow$ $\alpha $ & 27.7 $\pm$ 0.4 (4.2) & 5.3 $\pm$ 0.05 (0.5) \\ 
${}^{252}$Cf $\rightarrow$ $\alpha $ & 33.4 $\pm$ 0.6 (3.6) & 6.1 $\pm$ 0.09 (0.5) \\ 
${}^{252}$Cf $\rightarrow$ Long-range $\alpha $ & 64.2 $\pm$ 2.9 (23) & 9.4 $\pm$ 0.2 (2.1) \\ \hline 
\end{tabular}
\end{center}
\end{table}

\begin{table}
\caption{Mean values of ion ranges ($\mu $m); values of RMS are in parentheses.}
\begin{center}\begin{tabular}{c c} \hline 
${}^{84}$Kr  (3 \textit{A} MeV) & 34.4 $\pm$ 0.3 (2.4) \\ 
${}^{132}$Xe${}^{+26}$ (1.2 \textit{A} MeV) & 20 $\pm$ 0.1 (1.0) \\ 
${}^{86}$Kr${}^{+17}$ (1.2 \textit{A} MeV) & 17 $\pm$ 0.2 (1.0) \\ 
${}^{252}$Cf $\rightarrow$ 3 fragments & 5.1 $\pm$ 0.3 (2.0) \\ 
${}^{252}$Cf $\rightarrow$ 2 fragments & 9.1 $\pm$ 0.3 (0.6)\\ 
+ long-range $\alpha$ & 13.1 $\pm$ 0.6 (1.1) \\ \hline 
\end{tabular}
\end{center}
\end{table}

\vspace*{-8pt}
\section{Calibration by heavy ions}

NTE samples were exposed in the G. N. Flerov Laboratory of Nuclear Reactions, JINR at the IC-100 cyclotron to 1.2 \textit{A} MeV $^{86}$Kr$^{+17}$ and $^{132}$Xe$^{+26}$ ions and at the U-400M cyclotron to 3 \textit{A} MeV $^{84}$Kr ions [15]. The exposures were performed under vacuum conditions of the accelerators and also without a light protective paper. Fixing of the samples in the exposure chambers was performed at a light which is ordinary for a photo lab. For track observation the samples were installed with a significant inclination with the respect to the beam directions. Track densities in NTE reached 10$^{6}$ per cm$^{-2}$ for a few seconds of the sample exposures. Fig. 2 shows the range distribution of ions no scattered in NTE. Their average values are presented in Table 2. These data provide guidelines for further NTE calibrations at smaller energy values that are typical for the fission of heavy nuclei.

\begin{figure}
\centerline{\includegraphics*[width=5.81in, height=3.23in, keepaspectratio=false]{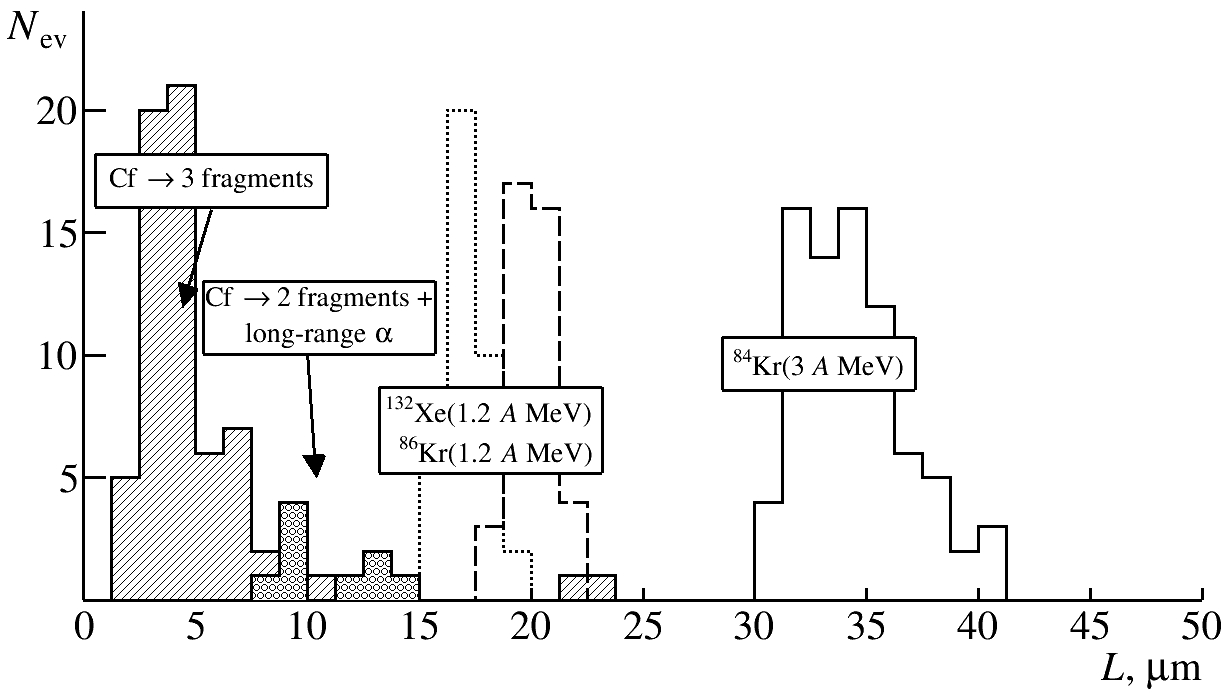}}
\caption{Distributions of ranges of ions ${}^{86}$Kr, ${}^{132}$Xe, ${}^{84}$Kr and in decays Cf $\rightarrow$ 3 fragments and Cf $\rightarrow$ 2 fragments $+$ long-range $\alpha $.}
\end{figure}

\begin{figure}
\centerline{\includegraphics*[width=5.93in, height=1.79in, keepaspectratio=false]{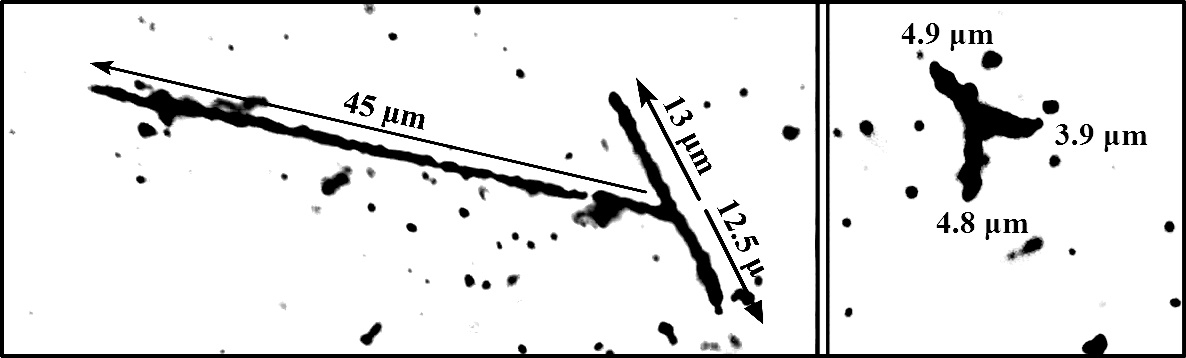}}
\caption{Examples of observed events of ternary fission; track lengths are specified. Left photo: long-range $\alpha $-particle (long arrow), fragment (middle arrow). Right photo: three fully observed fragment tracks.}
\end{figure}

\begin{figure}
\centerline{\includegraphics*[width=6.31in, height=1.99in, keepaspectratio=false]{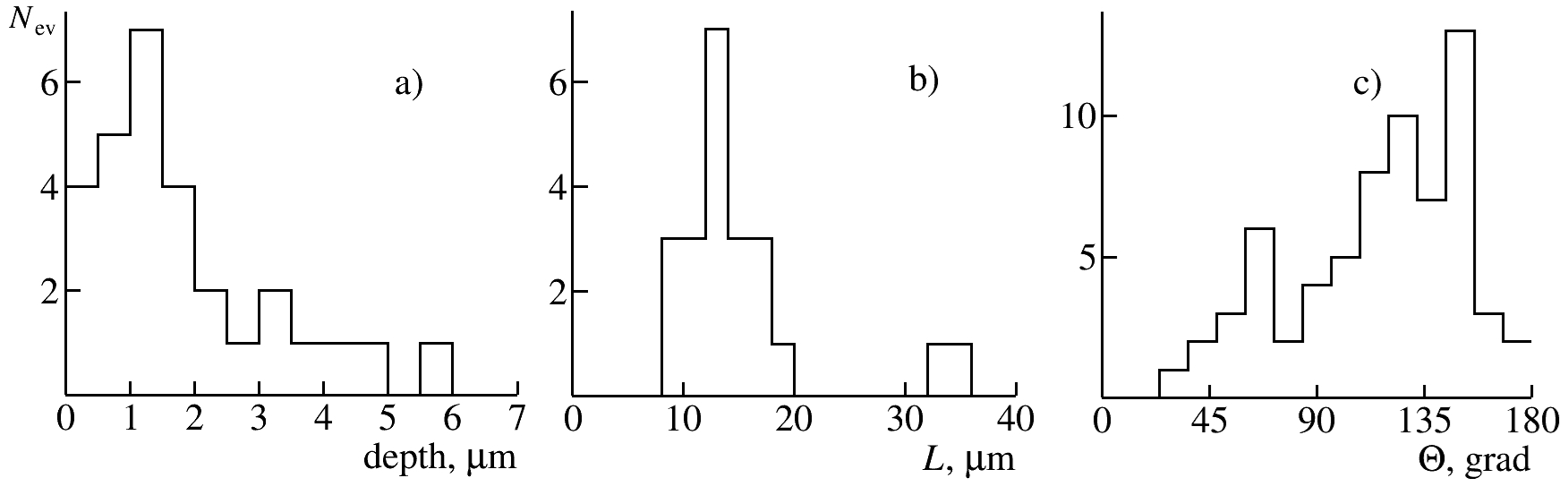}}
\caption{Distributions for the ${}^{252}$Cf fissions into three fragments: (a) over depth in NTE layer, (b) amount ranges of three fragments and (c) opening angles between fragments.}
\end{figure}

Search for heavy ion tracks on surface of the NTE samples exposed to the ${}^{252}$Cf source is carried out on the KSM microscope with a 15$\times$ objective. Usually MBI-9 microscopes are used for this stage. Using of KSM to search for very rare fission events has eased immediate transitions to their precise measurements with a 90$\times$ objective. Planar triples are found consisting of a pair of fragments and a long-range $\alpha $-particle as well as of fragments only. Their examples are given in Fig. 3. It is worth to emphasize a remarkable fact of the observation of triples in NTE but not only pairs of fragments. For such a full observation of triples their vertices should be dipped to a depth not less than a typical track thickness. Fig. 4a shows the distribution of vertices of Cf fissions into three fragments over NTE layer depth which has an average value of 1.8 $\pm$ 0.2 $\mu$m (RMS equal to 1.4 $\mu$m). Perhaps this effect is due to the binding of Cf atoms of AgBr micro crystals and their drift. Apparently, the source surface protection with initial thickness of the 50 $\mu$g/cm$^{2}$ gold deposition (according to the source passport) does not prevent such a penetration.

In 23 events of a true ternary fission, i. e., not containing $\alpha$-particles the ranges of all fragments (Fig. 3, right photo) are measured. Comparison of the mean values in Table 2 indicates that the average energy of fission fragments is of the order 400 \textit{A} keV. However, this is a very rough estimate. Calibration of the ranges of heavy ions should be promoted substantially below 1 \textit{A} MeV in controllable conditions provided by accelerators. Effective criteria for a fission into three heavy fragments is their amount range (Fig. 4b), which has an average value of 15.3 $\pm$ 1.4 $\mu$m when RMS 6.4 $\mu$m. In addition, the opening angles between the fragments are measured in these events (Fig. 4c). Their distribution is characterized by a mean value 116 $\pm$ 5$^{\circ}$ when RMS 36$^{\circ}$.

\vspace*{-8pt}
\section{Experience of automatic measurements}

The initial experience of a computer analysis of heavy ion tracks in NTE is obtained using the ImageJ program [16], available online and a close-up of the NTE sample exposed at an angle of 45$^{\circ}$ to the $^{132}$Xe$^{+26}$beam. Stages of such an analysis are shown in Fig. 5. initial close-up shot made by a NIKON camera D70 with 60$\times$ objective, track image finding, description of them as ellipses as well as determination of ion ranges in the computer (93 tracks) and results of manual measurements of non scattered tracks (40 tracks). In the computer case RMS is substantially greater than in manual one amounting 2.9 $\pm$ 0.2 $\mu $m and 1.0 $\pm$ 0.1 $\mu $m, respectively.

Often ion tracks entered in NTE are ending with bends or "forks" due to scattering on Ag and Br nuclei (Fig. 6). For example, in the case of ${}^{86}$Kr scatterings an average range to scattering points is equal to 7.7 $\pm$ 0.2 $\mu $m at RMS 1.8 $\mu $m which corresponds to an average energy at scattering of 250 $\pm$ 10 keV at RMS 100 keV. Residual tracks of scattered ions have a range 5.5 $\pm$ 0.3 $\mu $m at RMS 3 $\mu $m. It is impossible to attribute secondary tracks to an original ion and recoil target ion after scattering. Since only tracks of non scattered ions have been taken for the manual analysis, such a sampling has provided the perfect range resolution. 

Thus, the methodology prerequisites are established for transition to an automatic analysis on the discussed subject. The HSP-1000 microscope of DRD manufactured by "Seiko Precision" is unique scientific equipment at the European level. It is equipped with a high-resolution linear sensor, which allows up to 50 times higher speed of image acquisition compared to conventional CCD cameras. A full image of the sample is restored when continuous shooting from a relatively small number of long chains. Fully automatic digitization provides more easily and fast image analysis. This microscope with a 20$\times$ objective is used for the analysis of solid-state track detectors. In order to use the NTE resolution completely it is necessary to apply an oil immersed 60$\times$ objective. The HSP-1000 microscope is now being replenished in a corresponding way. 

Recently, the microscope HSP-1000 with a lens 20$\times$ was used to scan a large area of NTE exposed to 1.2 \textit{A} MeV $^{132}$Xe$^{+26}$ ions at a 45$^{0}$ inclination. An analysis by orders of magnitude beyond a man possibility is carried out on an array of 225 frames of 2500 $\times$ 2000 pixels. The program ImageJ has found 60,000 tracks on an area of 1.4 $cm^{2}$ and determined their lengths and planar angles $\Theta_{Xe}$ in an ellipse approximation (Fig. 7). The distribution of lengths is described by a Gaussian function with a parameter of 1.0 $\mu$m corresponding to the manual measurements. Statistics $\Theta_{Xe}$ is divided into two groups differing of 13$^{0}$ in the average values. These groups $\Theta_{Xe}$ are described by Gaussian functions with parameters of 3.4$^{\circ}$ and 2.0$^{\circ}$. Apparently, $\Theta_{Xe}$ splitting is caused by a change of a magnetic rigidity of approximately 20\% fraction of ions extracted from the accelerator IC-100 as a result of an electron pick-up in the residual magnetic field. The noticed effect can be used to analyze an ion beam composition. Currently, similar approaches to computer analysis are developed for the NTE exposures discussed above.
\vspace*{10pt}

\begin{figure}
\centerline{\includegraphics*[width=6.44in, height=4.61in, keepaspectratio=false]{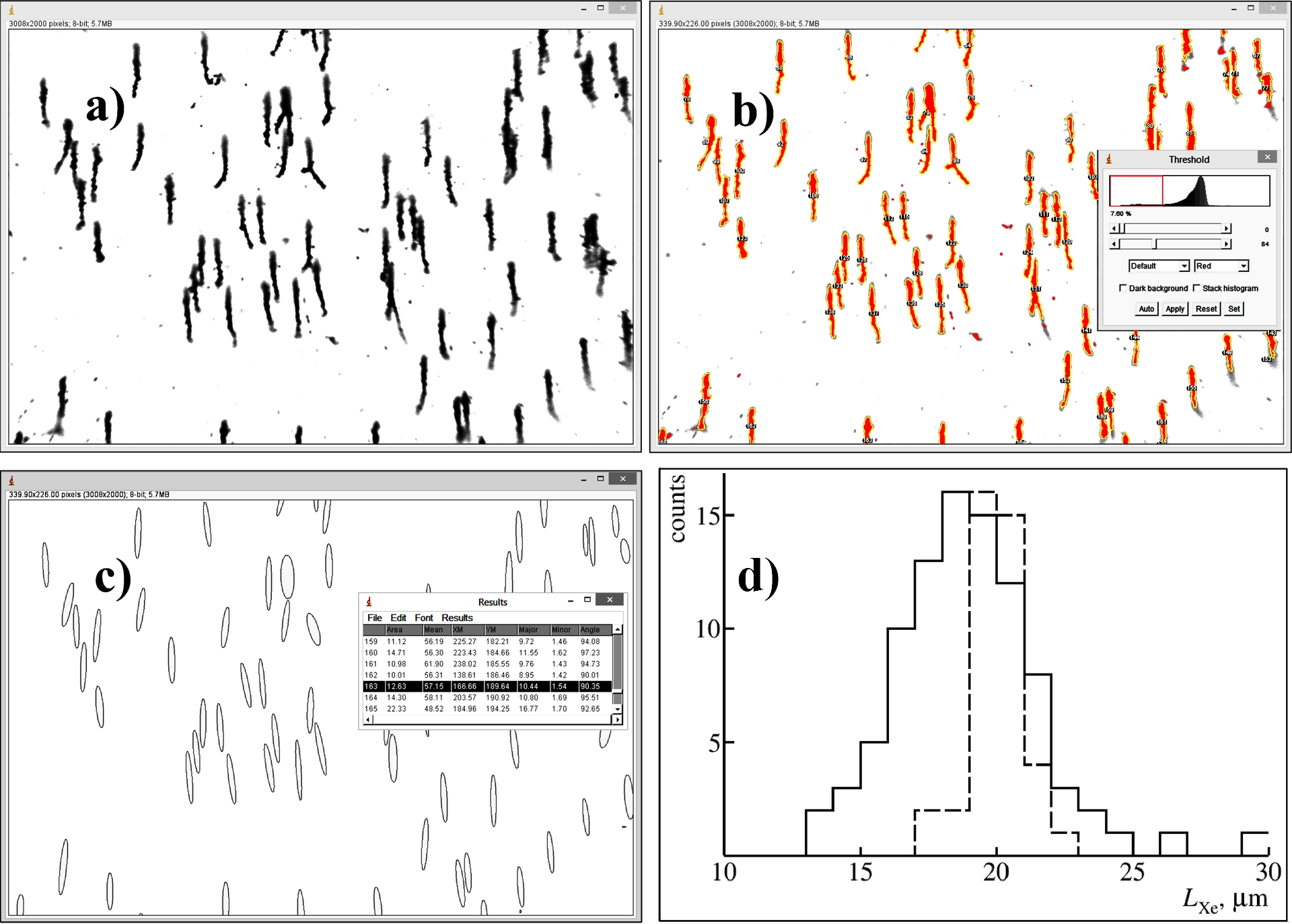}}
\caption{Stages of computer analysis: (a) initial close-up, (b) finding of track images, (c) description of them as ellipses and (d) ion range distribution in computer (solid line) and manual (dashed line) analysis.}
\end{figure}
\vspace*{-8pt}

\begin{figure}
\centerline{\includegraphics*[width=6.37in, height=2.72in, keepaspectratio=false]{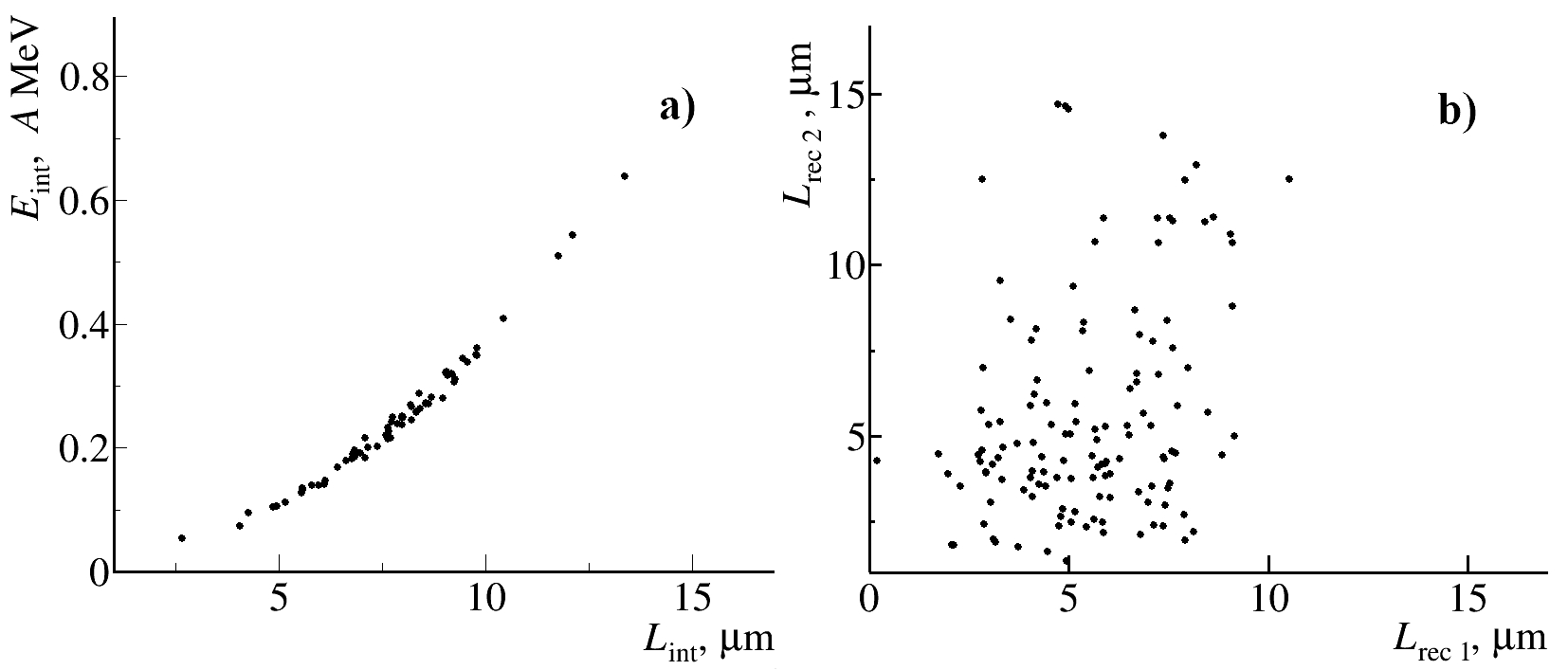}}
\caption{(a) Ranges of ions to scattering points \textit{L}${}_{int}$ and energy \textit{E}${}_{int}$ in them by the SRIM model; (b) ranges of recoil ions \textit{L}${}_{rec1}$ and \textit{L}${}_{rec2}$.}
\end{figure}
\vspace*{-8pt}

\begin{figure}
\centerline{\includegraphics*[width=4.47in, height=3.39in, keepaspectratio=false]{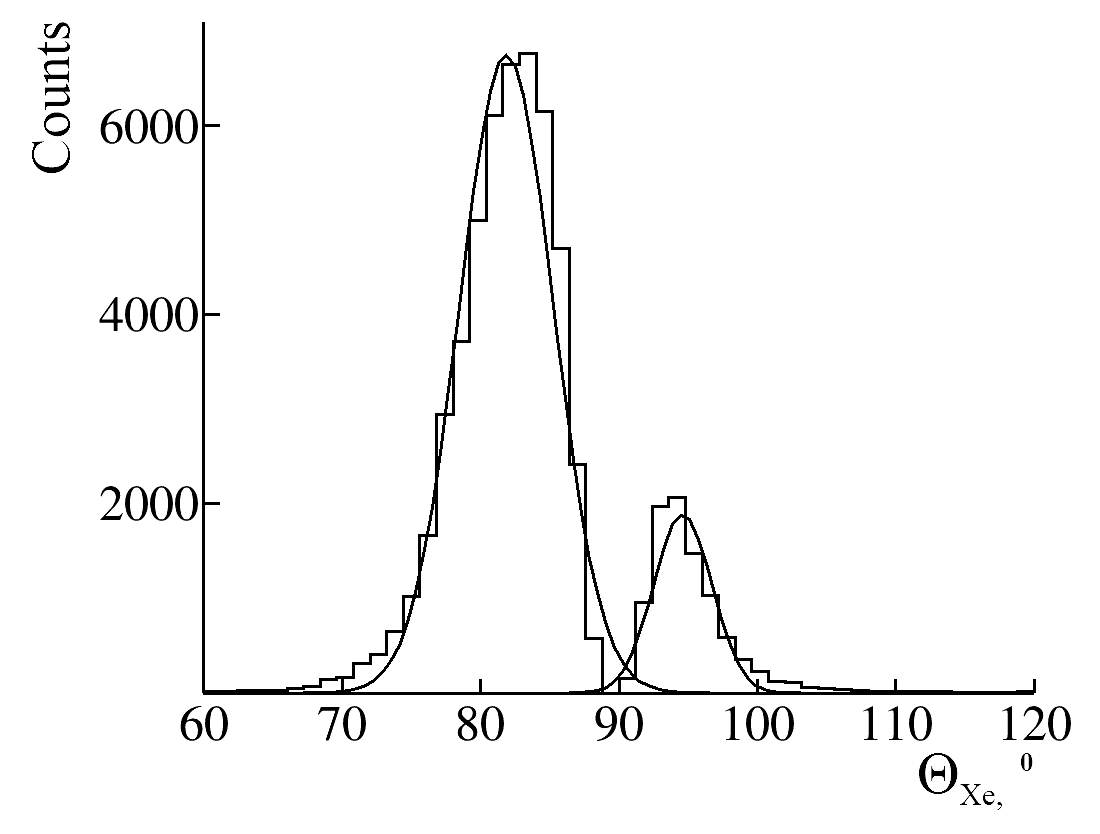}}
\caption{Distribution over planar angles of 1.2 \textit{A} MeV $^{132}$Xe ions.}
\end{figure}

\vspace*{-8pt}
\section{Conclusions}

The stated task of analyzing extremely rare events of the ternary fission is reduced to finding planar triples of nuclear fragments. Beginning at a common vertex and being randomly directed their tracks should have an extent of from 1 to 10 $\mu$m. Computer analysis of images will allow one to select the decays for a perfect manual analysis. Automation of search for ternary fission events will drastically reduce the most time-demanding stage and will help to focus the manual analysis on already found events. Thus, manual and automatic analyses complement each other. When the time of the NTE exposure is controlled the computer analysis can be applied in a desired scale and the diversity of tracks with estimation of ion energy, both to ion beam profilometry and to $\alpha$-dosimetry with random track directions.

In general, the synergy of modern radioactive sources, NTE proven metrology and advanced microscopy seems to be a promising prospect for $\alpha $-radioactivity and nuclear fission research. It can be assumed that ions of transfermium elements will be implanted some when in NTE. Their bright decays can be found as common vertices of few $\alpha $-particles and fission fragments. This perspective emphasizes the fundamental value of preservation and modernization of the NTE technique. Thus, the present study focused on the NTE return in practice of nuclear experiment will serve as a prototype of solution of an impressive variety of problems. Macro photos of the discussed exposures and videos based on them are available on the BECQUEREL project website [17].

\vspace*{-8pt}
\section{Acknowledgements}

The authors express gratitude to Profs. T. Hussein and M. Ghoneim (Cairo University), H.~El-Samman of the Minoufia University, Yu.~V.~Pyatkov (Moscow Engineering Physics Institute), A.~I.~Malakhov (JINR) and N.~G.~Polukhina (Lebedev Physics Institute FIAN) for discussions and support. This work was supported by grants of plenipotentiary representatives of governments of Bulgaria, Czech Republic, Egypt and Romania at JINR.




\begin{thebibliography}{}


\bibitem{powlers}Powell C.F., Fowler P.H., Perkins D.H. \textit{The Study of Elementary Particles by the Photographic Method}. London - New York: Pergamon Press. 1959.
\bibitem{Barkas}Barkas W.H. \textit{Nuclear Research Emulsions}. New York -- London: Academic Press. 1963

\bibitem{Gold}Goldschmidt-Cremont Y. Photographic Emulsions. \textit{Annu. Rev. Nucl. Sci.} 1953;141.
\bibitem{Kamanin}Kamanin D.V., Pyakov Y.V. Clusterization in ternary fission. ``\textit{Clusters in Nuclei''}. Lecture Notes in Physics; 2014;875:183.
\bibitem{Titterton}Titterton E.W., Brinkley T.A. Rare Modes in the Spontaneous Fission of Californium-252. \textit{Nature}. 1960;187:228.
\bibitem{Muga}Muga M.L., Bowman H.R. and Thompson S.G. Tripartition in the Spontaneous-Fission Decay of Cf$^{252}$. \textit{Phys. Rev.} 1961;121:270.
\bibitem{WebOdz}WEB site: \textit{http://www.odz.ujf.cas.cz/home/resources/microscope-hsp-1000.}
\bibitem{Slavich}``Slavich Company JSC''. WEB site: \textit{www.slavich.ru, www.newslavich.com.}
\bibitem{Kattabekov}Kattabekov R.R. \textit{et al}. Correlations of $\alpha$-particles in splitting of $^{12}$C nuclei by neutrons of energy of 14.1 MeV. \textit{Phys. At. Nucl.} 2013;76:add. issue (Russian)88-91. \textit{arXiv}: 1407.4575.
\bibitem{Artemenkov10}Artemenkov D.A. \textit{et al}. Exposure of Nuclear Track Emulsion to $^{8}$He Nuclei at the ACCULINNA Separator. \textit{Phys. Part. Nucl. Lett.} 2013;10:415-421. \textit{arXiv}: 1309.4808.
\bibitem{Zarubin}Zarubin P.I. \textit{et al}. $^{8}$He nuclei stopped in nuclear track emulsion. Proceedings of the 25th International Nuclear Physics Conference, held in Firenze (Italy), 2 -- 7 June, 2013. EPJ Web of Conferences. 2014;66:1-4. 
\bibitem{Arte12}Artemenkov D.A. \textit{et al}. $^{8}$He nuclei stopped in nuclear track emulsion. Proceedings of the 22nd European Conference on Few Body Problems in Physics, held in Krakow (Poland), 9 -- 13 September, 2014. \textit{Few-Body Systems}. 2014;55:733-736; \textit{arXiv}: 1410.5188.
\bibitem{Arte13}Artemenkov D.A. \textit{et al}. Exposure of nuclear track emulsion to thermal neutrons, heavy ions and muons.  Proceedings of the Conference on Fundamental interactions. MEPHI, Moscow, February, 2013. To be published in \textit{Phys. Atom. Nucl.}\textit{arXiv}: 1407.4572.
\bibitem{Ziegler}Ziegler J.F., Biersack J.P., Ziegler M.D. 2008. ``SRIM - The Stopping and Range of Ions in Matter''. ISBN 0-9654207-1-X., SRIM Co; WEB site: \textit{http://srim.org/.}
\bibitem{flerovLab}WEB site: \textit{http://flerovlab.jinr.ru/flnr/accelerators.html/.}
\bibitem{ImageJ}``Image processing and analyses in Java''. WEB site: \textit{http://rsb.info.nih.gov/ij/.}
\bibitem{BECQUEREL}``The BECQUEREL Project''. WEB site: \textit{http://becquerel.jinr.ru/.}
\end{thebibliography}
\end{document}